# Influence of swift heavy ion irradiation on structure and morphology of $La_{0.25}Pr_{0.375}Ca_{0.375}MnO_3$ perovskite film

Harsh Bhatt[1,2], Yogesh Kumar[3,*], R. B. Tokas[4], A.P. Singh[5], Fouran Singh[6] and Surendra Singh[1,2,*]

[1]Solid State Physics Division, Bhabha Atomic Research Centre, Mumbai 400085, India
[2]Homi Bhabha National Institute, Anushaktinagar, Mumbai 400094, India
[3]UGC-DAE Consortium for Scientific Research, R-5 Shed, BARC, Mumbai 400085, India
[4]Atomic and Molecular Physics Division, Bhabha Atomic Research Centre, Mumbai 400085, India
[5]5Physics Department, Dr. B. R. Ambedkar National Institute of Technology, Jalandhar 144027, India
[6]Material Science Group, Inter University Accelerator Centre, Aruna Asaf Ali Marg, New Delhi 110067, India

*lucky1708@gmail.com (YK), *surendra@barc.gov.in (SS)

Abstract:

The effects of $Ag^{15+}$ (120 MeV) swift heavy ion irradiation on the structural and morphological properties of epitaxial $La_{0.25}Pr_{0.375}Ca_{0.375}MnO_3$ (LPCMO) thin films was investigated by x-ray scattering and atomic force microscopy (AFM) techniques. LPCMO films of thickness ~ 280 Å were irradiated with an $Ag^{15+}$ ion beam at different fluences of $1 \times 10^{11}$, $5 \times 10^{11}$, and $1 \times 10^{12}$ ions/cm$^2$. XRD results suggested the development of the tensile stress along the out-of-plane direction of the LPCMO film upon ion irradiation which increased on increasing the ion fluence. The morphology of the film also modified with the irradiation and an increase in the fluence of the ion beam enhanced the in-plane height-height correlation length scale (grain size) with a loss of the fractal behaviour.

**Introduction:**

Manganite-based complex oxides are strongly correlated electronic systems which exhibit magnetoelectronic phase separation that plays a significant role in understanding the properties like colossal magnetoresistance (CMR), metal-to-insulator transition (MIT), colossal electroresistance (CER) and colossal piezoresistance (CPR) [1-5]. The competition and interplay among charge, spin, orbital, and lattice degrees of freedom lead to the coexistence of multiple phases and related electronic and magnetic properties in manganite oxides [1, 6]. Mixed valance bulk $(La_{1-y}Pr_y)_{1-x}Ca_xMnO_3$ (LPCMO) manganite, a prototypical electronic phase-separated system, has been studied extensively and it shows a phase separation of nanometer and micron length scale [5-11]. However thin films of LPCMO have exhibited properties distinct from bulk behaviour because of the additional handle (strain driven) available in the case of thin films suggesting that the strain field can tune the transport and magnetic properties as well as the phase coexistence [7, 12-20]. In the bulk phase, a nanometer and micron length scale phase separation has been observed for two manganites with slightly different compositions i.e. $(La_{1-y}Pr_y)_{1-x}Ca_xMnO_3$ ($y = 0.6$) with $x = 0.33$ and $x = 0.375$ [7, 9]. Different phenomena, like magnetic anisotropy driven single to multi-domain transition [12], electronic phase separation (EPS) [13-18], correlation of electronic and magnetic phases [16, 17], coupling of bending strain and magnetism [13,14,19,20], and electric field-driven phase separation [3, 21] have been reported for the thin film of LPCMO with ($x = 0.33$). However, the LPCMO film with $x = 0.375$ has been studied rarely even though a larger phase separation (length ~ micron) was suggested for this system in their bulk form [9].

Strain and related structural disorders can influence the competing interactions between different degrees of freedom and phase coexistence in the manganite films by modifying the structure and morphology. The effect of strain in manganite films is studied in both ways by applying strain (i) directly for substrate-induced strain [22-26], bending strain [4, 13,14,19,20], and (ii) indirectly i.e. ion implantation [27, 28] and swift heavy ion (SHI) irradiation [29-33]. The influence of SHI irradiation on the magnetic proximity effect in oxide-based manganite/superconductor heterostructure has also been demonstrated as a modification in structure, magnetic, and superconducting properties upon irradiation [34]. Recently, Matthews *et al.* [33] observed the formation and percolation of a network of disorder in the $LaMnO_3/SrTiO_3$ heterostructures on ion irradiation, which led to a change in structural and morphological properties at interfaces. Using x-ray resonant magnetic scattering (XRMS), Singh *et al.* [16, 17] correlated the electronic and magnetic phases across MIT temperature of the LPCMO thin film with $x = 0.375$ and found that these properties are highly dependent on the morphology of the interfaces. There are several ways to control the structure and morphology of the surfaces and interfaces of thin film and a suitable characterization in this respect is required to study its influence on macroscopic properties. Here, we report the evolution of structure and morphology upon ion irradiation of the LPCMO film with $x = 0.375$ and $y = 0.6$ (i.e. $La_{0.25}Pr_{0.375}Ca_{0.375}MnO_3$ film) grown on single crystal (110) $NdGaO_3$ (NGO) substrate. Combining the x-ray scattering and atomic force microscope (AFM) measurements we have studied the structural and morphological properties of the LPCMO film upon ion irradiation. These results suggest a strong modification in the structure and morphology of LPCMO film upon ion irradiation and these parameters are strongly correlated. The surface shows self-affine fractal behaviour which reduces upon irradiating the LPCMO film with a high ion dose.

**Experimental:**

The LPCMO thin film of thickness ∼280 Å was deposited from the high-density bulk targets on a single crystal (110) NGO substrate using a pulsed laser deposition technique [35]. A KrF excimer laser ($\lambda$ = 248 nm, pulse width = 20 ns) with an energy density of ∼4 J cm$^{-2}$ was used to ablate the high-density LPCMO target. The PLD chamber was evacuated to a pressure of $10^{-6}$ mbar before the deposition and during the deposition an oxygen partial pressure of 0.2 mbar and a substrate temperature of 750 °C were maintained. Annealing of the thin film was carried out at the same substrate temperature after deposition for 30 min in a 1000 mbar oxygen environment after which the film was allowed to cool naturally in the oxygen environment. The as-grown, 1cm x 1cm, LPCMO thin film was cut into four equal pieces and three pieces of the sample were irradiated at room temperature using 120 MeV $Ag^{15+}$ ions at a dose of $1 \times 10^{11}$, $5 \times 10^{11}$, and $1 \times 10^{12}$ ions/cm$^2$. Thus, for the present study, we have a pristine LPCMO film and three ion-irradiated LPCMO films with different doses mentioned above. The ion-irradiation experiment was carried out at Inter University Accelerator Centre (IUAC), New Delhi using the 15UD tandem accelerator. Structural characterization of the pristine and irradiated samples was carried out using x-ray scattering with Cu $K\alpha$ radiation from a rotating anode source, where x-ray diffraction (XRD) was used to investigate the atomic structure (strain) and x-ray reflectivity (XRR) was used to investigate depth-dependent structure. The surface topographic measurements were carried out using NT-MDT's solver-P47H ambient-based multimode AFM in intermittent-contact mode. An NT-MDT make silicon cantilever (model: NSG30) with a radius of curvature of cantilever tip < 10 nm, force constant of ∼30 N/m, and resonance frequency of oscillation ∼ 320 kHz was used for AFM measurements over a scan area of 10 μm × 10 μm and 5 μm × 5 μm with the resolution of 256 pixels × 256 pixels for each scan.

**Results and discussion:**

It is known that the defects induced by SHI irradiation provide an additional stress in the material and the case of thin films, the overall strain (due to substrate and irradiation) gets modified [29-33, 36-38]. Thus, it is believed that SHI irradiation can generate strain, release strain, anneal or create defects depending upon the initial state of the thin film. Various kinds of defects (columnar and point) can be produced in the material when an energetic ion is incident on it, which consequently modifies the local strain. There are mainly two processes of energy loss ($-dE/dx$) in the medium when the high energy ions pass through it [31] i.e., (i) nuclear energy loss, $S_n$ (elastic scattering) and (ii) electronic energy loss, $S_e$ (inelastic scattering) defined as: $-dE/dx = S_n + S_e$. However, loss in the energy for SHI is predominantly because of $S_e$. Using the Stopping and Range of Ions in Matter (SRIM-2013) code [39], we have simulated the energy loss for Ag ions in the LPCMO medium (Fig. 1(a)) as a function of the energy of the incident Ag ions. It is apparent from Fig. 1(a) that for 120 MeV $Ag^{15+}$ ions the value of $S_e$, $S_n$ and range in the LPCMO are 2.376 keV/Å, 0.011 keV/Å and 8.8 μm, respectively. Figure 1(b) shows the depth profiling of energy loss ($S_e$ and $S_n$) in the LPCMO/NGO thin film. This estimation of energy loss and range of ions in the LPCMO material suggests that Ag ions lose energy mostly for electron excitation and it will pass through the film (thickness ~ 280 Å << range) without being implanted. It is noted that the energy ($S_e$ = 2.376 keV/Å) deposited in the LPCMO layer on $Ag^{15+}$ ion irradiation is large enough to modify the properties of the film and using different doses of the ions will result in different structural and morphological properties.

The SHI irradiations induced structural modification were studied using XRD measurements. Figure 2(a) shows the XRD data for pristine LPCMO film as well as irradiated films with different doses. The magnified version of Bragg peaks around (220) and (330) reflection of NGO is shown in Fig. 2(b) and (c), respectively, for pristine and irradiated samples. The separation of Bragg peaks corresponding to the NGO substrate and film is near the (330) NGO reflection (corresponding 2θ ~ 73.5°) and thus it is used for estimation of the lattice parameter for the film. Inset of Fig. 2(a) shows the structure of the LPCMO film grown on orthorhombic NGO substrates along with the three mutually perpendicular directions of NGO substrate with (001) and (1$\bar{1}$0) as an in-plane direction and (110) as an out-of-plane (*c*) direction. The orthorhombic lattice of the NGO crystal can be represented as a distorted perovskite cube with its base inscribed in a slightly distorted square [40] i.e. the in-plane lattice parameters of NGO substrate are $d_{001}$ = 3.854 Å and $d_{1\bar{1}0}$ = 3.863 Å. Whereas the bulk LPCMO is a pseudo-cubic perovskite at room temperature with a lattice parameter ($d_{LPCMO}$) of 3.844 Å [41]. Therefore, different in-plane lattice parameters of the NGO substrates contribute to the tensile lattice mismatch strains in the two in-plane directions of LPCMO thin film, which can be calculated using $\varepsilon\ (\%) = 100 \times (d_{substrate} - d_{film})/d_{film}$ and are $\varepsilon_{001}$ = 0.26% and $\varepsilon_{1\bar{1}0}$ = 0.49%. This suggests that the $c$ lattice parameter for LPCMO film on (110) NGO substrate should compress due to this lattice mismatch. However, XRD data from pristine film suggested a small increase in the $c$ lattice parameter of LPCMO film due to lattice mismatch. Earlier studies [3,4,12] also suggested similar results where a perfect match for the $c$ lattice parameter for LPCMO film with NGO was observed, however, the lattice mismatch strain highly depends on the thickness of the film. XRD data from ion irradiated films show a systematic shift in the Bragg peak positions to a lower angle except at the first ion irradiation with a fluence of 1×10$^{11}$ ions/cm$^2$. The shift in the Bragg peaks on ion irradiation of the LPCMO film was used to estimate the $c$ lattice parameter of the film and plotted in Fig. 2 (d) along with the bulk value of lattice constant for (110) NGO and LPCMO (horizontal lines in Fig. 2 (d)). It is evident from Fig. 2 (d) that the $c$ lattice parameter increases on irradiation for the higher fluences of 5×10$^{11}$ and 1×10$^{12}$ ions/cm$^2$ with the $c$ lattice parameter of 3.936 Å for a fluence of 1×10$^{12}$ ions/cm$^2$. This change in lattice parameter suggests the advancement of a strong tensile strain along the normal of the film upon ion irradiation with a fluence > 10$^{11}$ ions/cm$^2$.

To study the influence of ion irradiation on the depth profiling structure of LPCMO film, we have carried out specular (angle of incidence = angle of reflection) XRR measurements. Experimental specular XRR data (symbols) [42, 43] with corresponding fits (solid lines) as a function of wave-vector transfer, $Q = 4\pi\ (\sin \theta)/\lambda$ (where $\lambda$ is the wavelength of x-ray and $\theta$ is the angle of incidence) for pristine and irradiated LPCMO films, which are sifted vertically for better visualization, are shown in Fig. 3 (a). XRR is qualitatively related to the Fourier transform of the electron scattering length density (ESLD) depth profile, [42, 43] averaged over the whole sample area. In addition, we have performed XRR measurements from different parts of the pristine sample using small beam sizes and observed similar XRR profiles, confirming that the sample is uniformly grown. The parameters of the ESLD depth profile also included layer thickness and interface (or surface) roughness. The best-fitted reflectivity curves for different samples are shown as a solid line in Fig. 3(a) and the corresponding ESLD depth profiles are shown in Fig. 3 (b) to (e). Different structural parameters obtained from XRR data for pristine and irradiated LPCMO films are given in Table 1. The thickness of pristine LPCMO film was found to be ~ 275 Å with a surface roughness of ~ 9 Å. XRR results from pristine LPCMO film also suggested a small increase in the ESLD value towards the surface region. Upon first ion irradiation with a fluence of $1\times10^{11}$ ions/cm$^2$, we find a small increase (~ 2.3%) in the ESLD and a drastic reduction in the surface roughness of the LPCMO film. This may be a result of the annealing of the film upon ion irradiation at a lower fluence value [37]. However, a decrease in ESLD was observed on irradiating the LPCMO film with higher ion fluence. Therefore, XRR in combination with XRD suggests a significant modification in structural properties both at layer structure length (with sub-nanometer depth resolution) and atomic (XRD) length scale due to SHI irradiation.

Further to investigate the irradiation-induced morphology in the LPCMO film we carried out AFM measurements for the pristine and irradiated films. Figure 4 shows the AFM results from the pristine and irradiated samples. AFM is an advantageous tool to extract the microstructure/microroughness/morphological information of surfaces with high vertical and spatial resolution [44]. Although the root mean square (RMS) roughness estimated from the AFM technique is highly local and dependent on the scan area, it provides a complete height-height morphology for describing the fractal geometry and scaling concepts [45-47], which are important factors for influencing the interactions and macroscopic properties of a thin film. Column (i) of Fig. 4(a) to (d) shows the 3D-AFM images of pristine and irradiated LPCMO films recorded over a scan area of 5.0 μm × 5.0 μm. Visually, AFM images indicate a modification in the morphology of the film upon ion irradiation and suggest a distribution in LPCMO islands in both out-of-plane and in-plane directions. Therefore a comparison of the surface topography of pristine and irradiated LPCMO films suggests the presence of grains of similar sizes on the surface, which are distributed densely over the substrate for the pristine film. Whereas the topography of the irradiated film depicts grains of different sizes, which are larger in size compared to the pristine film. More qualitative information has been obtained from the power spectral density (PSD) function of these surfaces and their characteristic parameters described later. Further to analyze the distribution of these islands (grains), we have plotted the corresponding topographical histogram (symbols) and the Gaussian fit (solid black lines) to histogram data for pristine and irradiated films in the second column (ii) of Fig. 4(a) to (d). A single Gaussian function fits well to the histogram profile for pristine film suggesting that the surface has features distributed symmetrically around a mean surface profile. While the single Gaussian function also fits the histogram for irradiated LPCMO film with the lowest ions fluence of $1\times10^{11}$ ions/cm$^2$, the increase in ion fluence modifies the height-height distribution, and the deviation of single Gaussian fit for irradiated films indicates that the surface features are distributed around more than one mean surface profiles.

The surface morphology recorded by AFM can be analyzed by considering either the height-height autocorrelation function, $C(r) = \langle Z(r)Z(0) \rangle$, where Z denotes the height at the interface/surface and $r$ is the spatial distance between two points ($x1,y1$) and ($x2, y2$) on the surface; or a height difference correlation function, $g(r) = \langle [Z(x2,y2) - Z(x1,y1)]^2 \rangle$ [45]. For self-affine fractal surfaces, these two correlation functions are related as [45]: $g(r) = 2\sigma^2 - 2C(r) = 2\sigma^2[1 - \exp\{-(r/\xi)^{2h}\}]$, where $\sigma$ is the root mean square of the surface roughness, $\xi$ is the correlation (cutoff) length (i.e. a measure of the lateral length scale of roughness) and $h$ is the Hurst parameter or roughness exponent and it describes the texture of roughness of the interface/surface. The Hurst parameter $h$ can take a value $0 < h < 1$ and for a self-affine morphology it defines the fractal dimension $D = 3 - h$ of the interface. Column (iii) of Fig. 4 (a) to (d) show the experimental $g(r)$ data (symbol) extracted from height-height information from the AFM image of scan area of 5.0 µm × 5.0 µm and the corresponding fits (solid black line) assuming self-affine surface for pristine and irradiated LPCMO films. We find small changes in morphological parameters ($\sigma$, $\xi$ and $h$) on irradiation of films with different doses and parameters $\sigma$, $\xi$ and $h$ showed a variation in the range of 13±5 Å, 2500±400 and 0.90±0.05, respectively. Thus using these height-height correlation functions, the surface morphology for all the films indicates a fractal dimension of ~2.1.

While the self-affine fractal nature studied using the height-height difference function suggested modification in surface morphology upon ion irradiation, we further studied the evolution in surface topography by the PSD function [45-47] based on the fast Fourier transform (FFT) to determine the dominant spatial frequencies in a surface of LPCMO film upon ion irradiation. We have calculated PSD spectrums as a function of spatial frequency from AFM images of pristine and irradiated samples, shown as open circles in column (iv) of Fig. 4 (a) to (d), using the method discussed elsewhere [46]. The PSD data for different samples clearly show a variation over the frequency ranges and thus need a careful analysis for the interpretation of data over the entire frequency range. PSD of the surface topography is a fundamental tool for describing the statistical properties of the surfaces and analyzing an AFM image into waves with spatial frequencies by Fourier transforms. However, to interpret and understand the PSD information quantitatively more appropriate analytical models are required and there are several models used to explain the data in the different frequency ranges [46, 47]. We have fitted the PSD data using a combination of three models which define PSD data in different frequency ranges [46, 47] and are (i) inverse power law, (ii) the K-correlation or ABC model, and (iii) the Gaussian function.

The self-affine fractal surfaces follow the inverse power law, i.e. $PSD_{fractal} = K/f^v$, where $K, f,$ and $v$ are the spectral strength of fractal, spatial frequency, and fractal spectral indices, respectively and thus it is one of the standard models to fit PSD data for this contribution. For self-affine fractal surfaces this contribution is calculated as the slope of logarithmic PSD versus $f$ plot at the high spatial frequencies and depending upon the value of the spectral indices ($v$) one can estimate the fractal dimension "$D$" [47]. Since this model (fractal) cannot provide information on the physical roughness parameter, the $K$-correlation (ABC model) model is used to describe random rough surface morphology and it is defined as [46, 47]: $PSD_{ABC} = \frac{A}{(1+B^2f^2)^{\frac{C+1}{2}}}$, where A, B, and C are model parameters and depend on the specific surface structure [46, 47]. These two models, which are monotonically decreasing functions of spatial frequency, generally provide a full description of surface morphology from PSD data. However, the additional features on surfaces, like superstructures, etc., provide additional contributions (local maxima) in PSD data in the lower frequencies and thus Gaussian functions are used to describe such features in the PSD curve. The Gaussian contribution to PSD is defined as [46]: $PSD_{Gauss} = \sum_i \pi\sigma_{sh,i}^2\tau_{sh,i}^2 \exp[-\pi^2\tau_{sh,i}^2(f-f_{sh,i})^2]$, where $\tau_{sh}$ and $\sigma_{sh}$ corresponds to the size and height of the superstructures on the surface and summation is used for considering more than one length scale for superstructures [46]. We have used these three contributions for the PSD model and fitted the experimental PSD data and also assumed the summation of two Gaussian functions to get the best fit to PSD data. Different contributions to the PSD model and best fit are also shown in Fig 4 (column (iv)). The PSD model characteristics parameters obtained from the fit to PSD data for pristine and irradiated films are given in Table 2. The PSDs of the LPCMO films under different conditions (pristine and irradiated) exhibit inverse power law variation at the high spatial frequency region confirming the presence of fractal components in the surface topographies. The spectral indices ($v$) for all the films, which is a measure of slopes of the PSD curve at the high-frequency region, are almost the same (see Table 1), with a value of 2.1 to 2.25. Using a relation for estimating the fractal dimension from spectral indices [44] i.e. $D = \frac{7-v}{2}$ $for$ $1 \leq v \leq 3$, we obtained a fractal dimension of ~ 2.4 (for $v = 2$), which is close to the value we have estimated ($D$ ~ 2.1) from the height-height difference function for self-affine surfaces. In contrast, the spectral strength (*K*) decreases on irradiation of film indicating that upon ion irradiation the film loses the fractal behavior.

A comparison of experimental PSD profiles for pristine and irradiated LPCMO films is shown in Fig. 5 (a). Inset shows the PSD function in the higher spatial frequency region. It is evident from Fig. 5 (a) that a drastic change in the PSD profile over the whole frequency range is observed on the first ion irradiation of the film by an ion dose of $1 \times 10^{11}$ ions/cm$^2$, this is consistent with the results from XRR measurements, which showed densification and reduction of the layer structure. The reduction in PSD profiles in the intermediate frequency range for the irradiated sample at low fluence also suggests a reduction in the roughness parameter for this sample. While the fractal behaviour of these films shows similar behaviour but the slower or faster variation of PSD over the spatial frequency suggests different topological behaviour on ion irradiation. On first ion irradiation of the film with a fluence of $1 \times 10^{11}$ ions/cm$^2$, we observed that PSD falls faster in higher frequency regions as compared to other conditions (pristine and other fluence) suggesting modification in lateral structure (grain sizes). On increasing the dose of ion irradiation in the LPCMO film, we observed slower variation in the PSD profiles, indicating the presence of a larger in-plane length scale (grain size) in the system. A similar variation of PSD profiles in the low and intermediate frequency range suggests the same roughness parameters for the films. However, the intrinsic roughness parameters, like RMS roughness ($\sigma_{ABC}$) and lateral correlation length ($L_{ABC}$), can be estimated from the *K*-correlation or ABC model of PSD, as mentioned above, using [46, 47]: $\sigma_{ABC}^2 = 2\pi A/B^2(C-1)$ and $L_{ABC}^2 = (C-1)^2 B^2/2\pi^2 C$. We have calculated these characteristic parameters from the PSD data and plotted them in Fig. 5 (b) for pristine and irradiated LPCMO films. We find a small decrease in RMS roughness and correlation length of the first irradiated LPCMO film at a fluence of $1 \times 10^{11}$ ions/cm$^2$ as compared to pristine film and a further increase in ion fluence increases both these parameters.

Figure 5 (c) depicts the comparison of lattice parameter and fractal dimension for LPCMO films under different conditions (pristine and ion irradiation at different fluence). Figures 5 (b) and (c) illustrate the correlation between atomic structure and morphological properties of LPCMO films under ion irradiation with different fluences. A similar variation of RMS roughness, lateral correlation length, and out-of-plane lattice parameters (*c*) as a function of ion fluence clearly suggests a strong correlation in structural and morphological modifications due to ion irradiation. The study also suggests an increase in in-plane grain size and a reduction in fractal characteristics of the film surface on ion irradiation with a fluence of larger than $1\times10^{11}$ ions/cm$^2$.

**Conclusion:**

In summary, we have studied the effect of SHI ion irradiation on the structure and surface morphology of LPCMO thin films deposited on NGO substrate. XRD measurements from irradiated LPCMO suggested an increase in the out-of-plane lattice parameter on increasing the dose of ions. AFM measurements were analyzed mainly through the PSD function. The SHI-irradiated LPCMO thin films display a self-affine surface. Due to irradiation, the surface structure and morphology of the films are modified significantly. A comparison of structural and morphological parameters as a function of ion dose suggested a strong correlation in atomic structure and morphology, where the lattice constant (fractal behaviour) increases (decreases) on increasing the ion dose of irradiation.

**Acknowledgments:**

One of the author (YK) would like to thank the Department of Science and Technology (DST), India, for financial support via the DST INSPIRE faculty research Grant (No. DST/INSPIRE/04/2015/002938) and the Science and Engineering Research Board (SERB), India via research grant (No. SB/SRS/2021-22/65/PS).

Table 1: Structural parameters obtained from X-ray scattering measurements.

| Samples → | Pristine | Irradiated with 120 MeV $Ag^{15+}$ ions with different doses | | |
|---|---|---|---|---|
| | | $1\times10^{11}$ ions/cm$^2$ | $5\times10^{11}$ ions/cm$^2$ | $1\times10^{12}$ ions/cm$^2$ |
| Thickness (Å) | 275±5 | 245±11 | 265±7 | 270±5 |
| Roughness (Å) | 9±3 | 3±2 | 3±2 | 4±2 |
| ESLD ($10^{-5}$ Å$^{-2}$) | 4.34±0.15 | 4.44±0.18 | 4.22±0.15 | 4.15±0.14 |
| $c$ lattice parameter (Å) | 3.86 ±0.01 | 3.86±0.01 | 3.90±0.01 | 3.93±0.01 |

Table 2: Different parameters describing a PSD model are fitted to experimental PSD data for pristine and irradiated LPCMO films.

| Samples (LPCMO film) | Fractal | | k-Correlation | | | Shifted Gaussians | | | | | |
|---|---|---|---|---|---|---|---|---|---|---|---|
| | | | | | | Guassian1 | | | Gaussian2 | | |
| | $K$ ($10^{-3}$ nm) | $v$ | $A$ ($10^4$ nm) | $B$ (nm) | $C$ | $\sigma 1$ (nm) | $\tau 1$ (nm) | $x1$ (μm$^{-1}$) | $\sigma 2$ (nm) | $\tau 2$ (nm) | $x2$ (μm$^{-1}$) |
| Pristine | 15 | 2.1 | 15 | 425 | 7 | 1.0 | 3200 | 0.8 | 0.8 | 1100 | 3 |
| $1\times10^{11}$ ions/cm$^2$ | 13 | 2.15 | 9.2 | 400 | 7.25 | 1.0 | 3500 | 0.7 | .7 | 1000 | 1.5 |
| $5\times10^{11}$ ions/cm$^2$ | 2 | 2.2 | 25 | 500 | 7 | 1.2 | 3500 | 0.7 | 1.2 | 700 | 1.5 |
| $1\times10^{12}$ ions/cm$^2$ | 2 | 2.25 | 27 | 510 | 7 | 1.3 | 3000 | 0.7 | 1.1 | 700 | 1.5 |

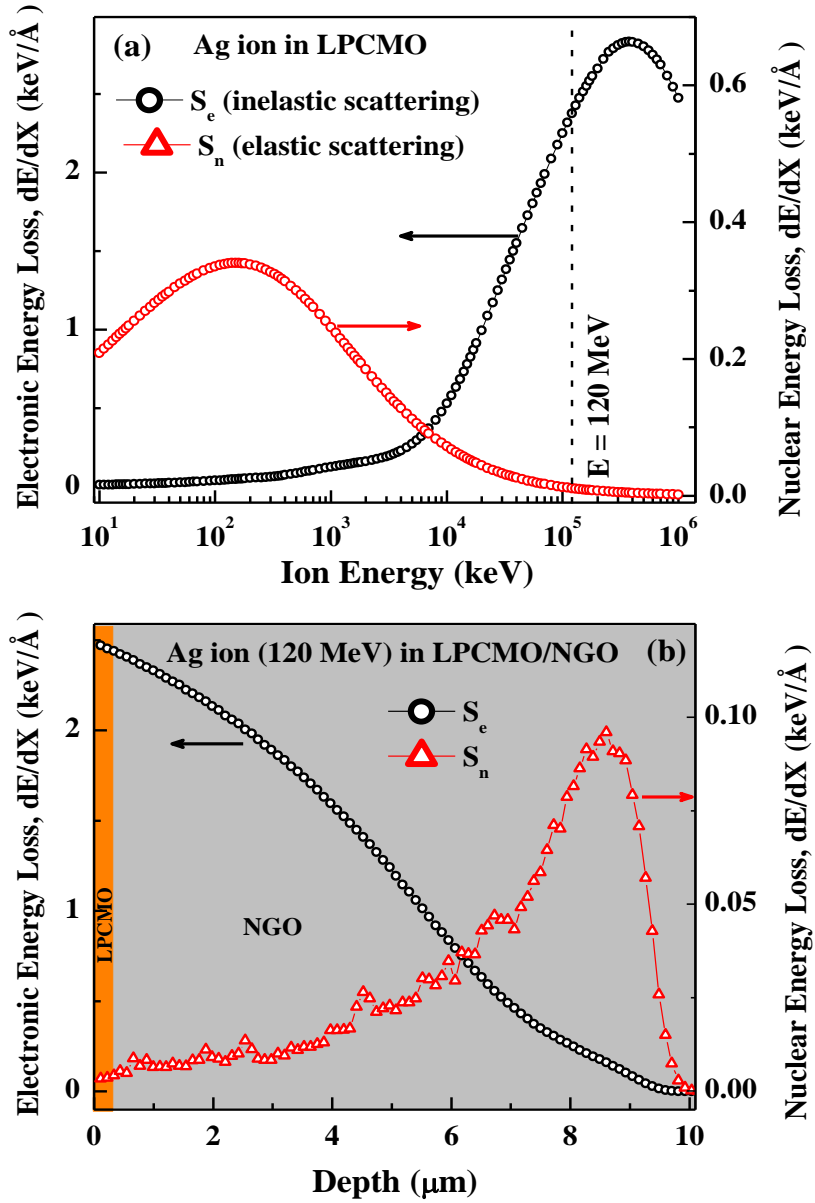

Figure 1: (a) Energy loss (electronic and nuclear contribution) of $Ag^{15+}$ ions in LPCMO as a function of the energy of incident ions. (b) Depth-dependent nuclear and electronic energy loss profile of 120 MeV $Ag^{15+}$ ions in an LPCMO (275 Å)/NGO film.

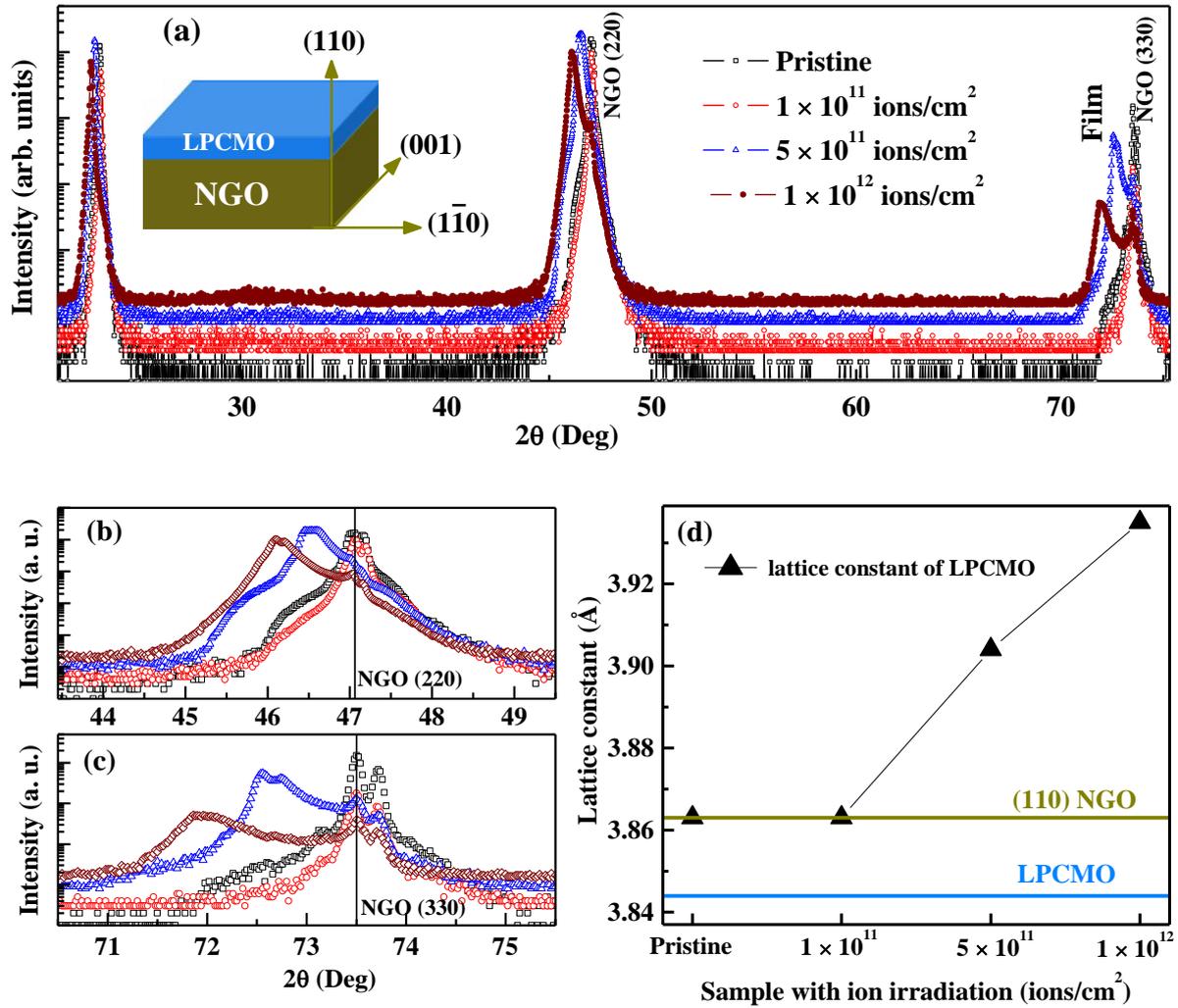

Figure 2: XRD pattern for pristine LPCMO film grown on NGO substrate and ion irradiated LPCMO films at different doses of $1 \times 10^{11}$, $5 \times 10^{11}$, and $1 \times 10^{12}$ ions/cm². Inset shows the schematic of the structure of the LPCMO film on NGO substrate with three mutually perpendicular directions of NGO substrate. The magnified version of the XRD pattern near (b) (220) and (c) (330) Bragg peaks of NGO substrate, highlighting the shift of Bragg peak for a film to a lower angle on increasing the dose of ion irradiation. (d) estimated out-of-plane, *c*, the lattice parameter of the film at different ion doses.

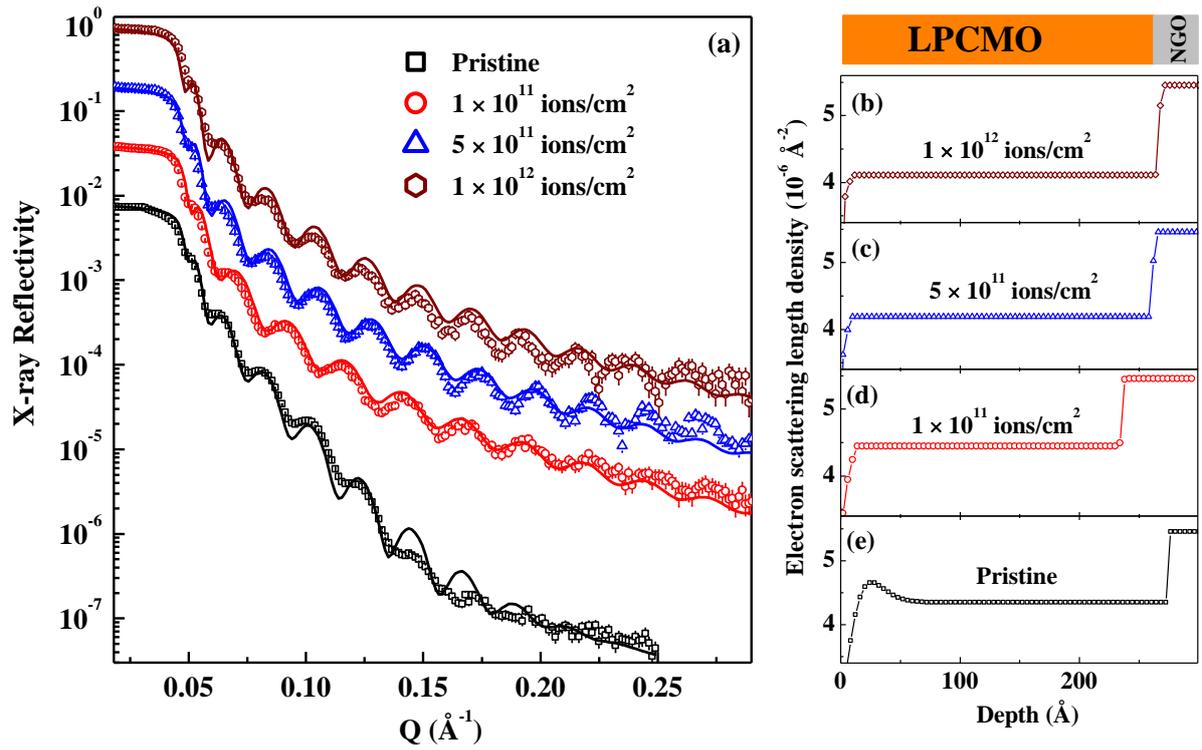

Figure 3: (a) XRR data (symbol) and corresponding fits (solid lines) for pristine and irradiated LPCMO films, which are shifted vertically for better visualization. (b) to (e): Electron scattering length density depth profiles obtained from XRR data for pristine and irradiated samples at different doses.

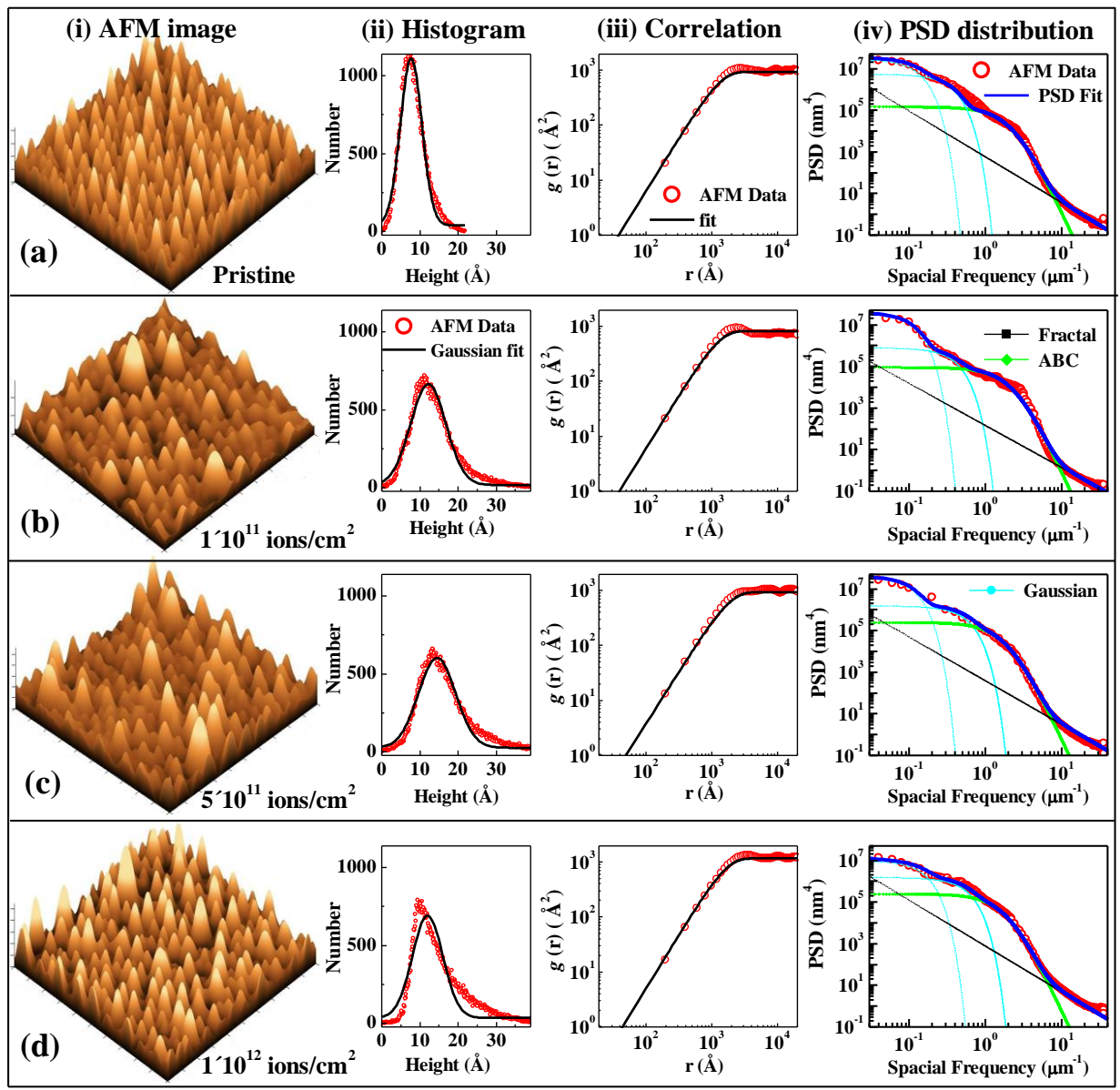

Figure 4: Comparison of AFM measurements from (a) pristine and ion irradiated LPCMO films at different doses of (b) $1 \times 10^{11}$ ions/cm$^2$, (c) $5 \times 10^{11}$ ions/cm$^2$, and (d) $1 \times 10^{12}$ ions/cm$^2$. Columns (i), (ii), (iii), and (iv) in each case (a to d) show the 3D-AFM images of size 5μm × 5μm, histogram, height-height correlation function, and PSD distribution function, respectively for pristine and irradiated LPCMO films.

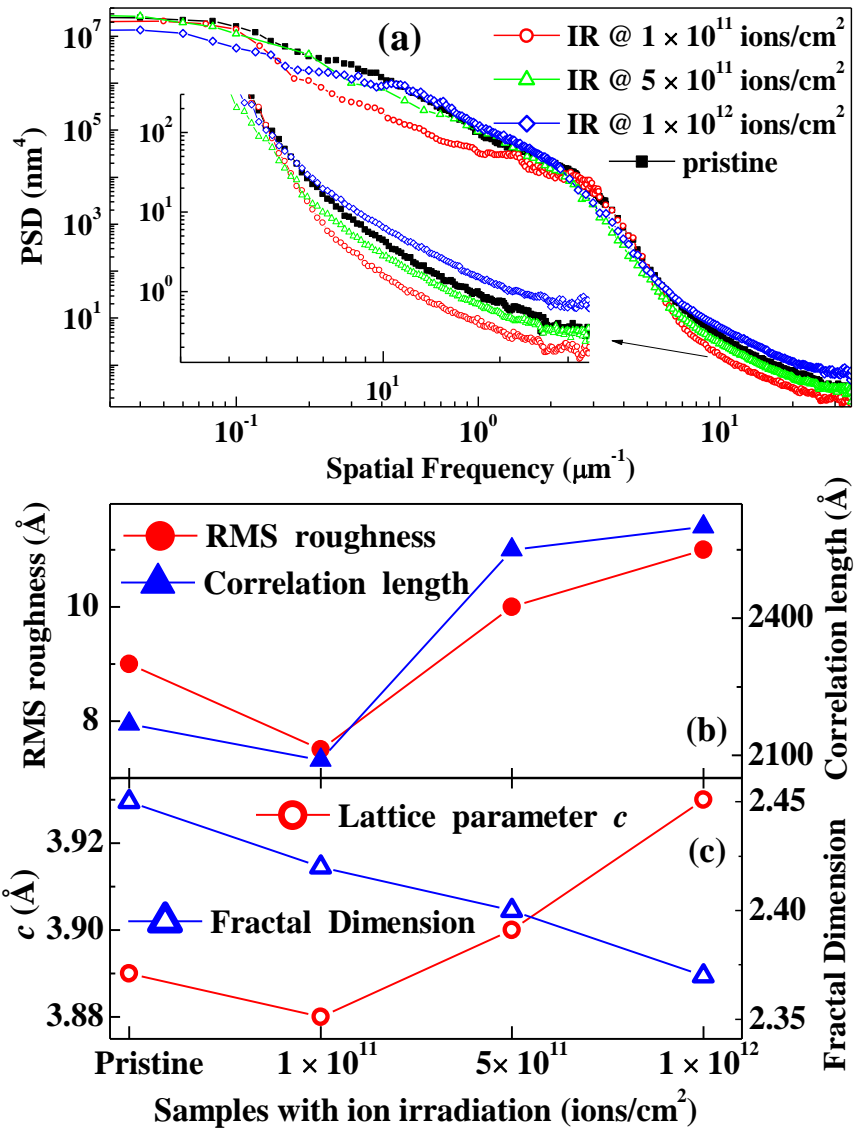

Figure 5: (a) The experimental PSD profiles of LPCMO film under different conditions of ion irradiations. Inset shows the magnified version of the PSD profile for a spatial frequency range of $f > 3$ μm$^{-1}$. Evolution of (b) RMS roughness and lateral correlation length, and (c) lattice parameter and fractal dimension, upon ion irradiation with different fluence.